\documentclass[a4paper]{article}
\usepackage{cite}
\usepackage{amsmath,amssymb,amsfonts}
\usepackage{algorithmic}
\usepackage{graphicx}
\usepackage{textcomp}
\usepackage{xcolor}
\usepackage{comment}
\begin{document}

\title{Earthquake detection at the edge: IoT crowdsensing network}
\date{}

\author{Enrico Bassetti, Emanuele Panizzi \\
\textit{Department of Computer Science} \\
\textit{Sapienza University of Rome}\\
Rome, Italy \\
\{bassetti, panizzi\}@di.uniroma1.it
}

\maketitle

\begin{abstract}
Earthquake Early Warning state of the art systems rely on a network of sensors connected to a fusion center in a client-server paradigm. Instead, we propose moving computation to the edge, with detector nodes that probe the environment and process information from nearby probes to detect earthquakes locally. Our approach tolerates multiple node faults and partial network disruption and keeps all data locally, enhancing privacy. This paper describes our proposal's rationale and explains its architecture. We then present an implementation using Raspberry, NodeMCU, and the Crowdquake machine learning model.
\end{abstract}

\section{Introduction}

Many countries perform earthquake detection through a national network composed of hundreds of high-precision seismic stations. Each seismometer in a station has high sensitivity and can perceive low magnitude or very distant earthquakes (sometimes from other countries). By interpolating signals from three or more seismic stations, it is possible to localize the epicenter and compute the magnitude.
These seismic networks are costly, and building them might be a decades-long process. Some countries use such networks to provide an Earthquake Early Warning system, like the Japanese one by JME\cite{hoshiba2008earthquake}. 

An alternative that has been getting traction in the last decade is the crowdsensing EEW network, based on the availability of low-cost MEMS sensors together with the widespread internet connection. Volunteers can participate in crowdsensing using their smartphone or an IoT sensor as a seismometer. 
Crowdsensing EEW tackles the problem of MEMS low precision by trading quality with quantity. By leveraging the lower cost of smart devices and distributing such costs among participants, these systems have a large user base and thus many seismometers, i.e., thousands or more. This approach has proven to be successful, for example in \cite{allen2020myshake}, at least to estimate the epicentral area and an approximated intensity.

Existing crowdsensing EEW networks adopt a centralized processing approach: seismometers send the collected data to a \emph{fusion center} that processes it to understand whether the report is a quake signal or not. In some cases, the sensors send MEMS raw signal to the fusion center (dumb approach). Other times, the edge sensors do partial calculations (limited due to resource constraints) and send preprocessed data. The fusion center does the detection work adopting a post-processing filtering that involves signals from many local seismometers to exclude false positive or false negative earthquake detections. 

As explained in the next section, this kind of architecture has some drawbacks that motivate our work.

In this work, we propose a peer-to-peer distributed EEW architecture based on edge computing. Each network node can sense the environment and detect a local earthquake without relying on a fusion center or a leader node. It can share this information with its neighbors, propagating the detection. This system keeps all data locally and can tolerate multiple node faults and some partial network disruption.

\section{Motivation}

Ideally, an EEW system should be fault-tolerant, which means that if one or a few of its components fail, the overall system can continue to work seamlessly. In the absence of fault tolerance, the High Availability property (HA) might be helpful: HA systems tolerate a stop or downtimes between the fault and its recovery. However, if a fault occurs during an earthquake, the EEW system might be unavailable at a critical moment. 

EEW systems currently built or proposed in the literature do not have a fault-tolerant architecture, as the fusion center constitutes a Single-Point-of-Failure (SPOF) that, if unavailable, prevents the entire system from working. We should also consider the connections to the fusion center, like international internet links with sensors, as a system component that can fail, causing the isolation of the fusion center and thus the unavailability of the EEW system.

Nevertheless, the fusion center can be HA by implementing it as a distributed system in cloud providers. However, this still creates a SPOF.

The first motivation behind our proposal is to solve the availability problem. As we describe in Section \ref{sec:fault}, our system can tolerate multiple node faults and some partial network disruption. 

The mainstream EEW architecture also has a privacy-related issue. It is possible to process raw accelerometric data to extract information other than seismic. For example, it is possible to detect some spoken words using the accelerometer in place of the microphone \cite{zhang2015accelword}. As another example, we experienced in our work that we could correlate the noise level of seismometers in our homes with the presence of people at home. So, sending raw seismic signals to a fusion center might expose them to unwanted processing that can violate the users' privacy: an attacker could discover information about a family's life habits or even extract words of private conversations.
Our proposed architecture enhances the privacy of the detection, keeping locally the sensitive data collected in private places by a crowdsensing EEW system.

\section{Related work}

Decentralized approaches to earthquake detection have been studied for years. Tsitsiklis \cite{tsitsiklis1990decentralized} proposed a decentralized detection architecture where a central system (named ``fusion center'') collects ``messages'' from sensors. In EEW, a ``message'' can be a signal sample that the sensor claims to be a quake signal. Faulkner et al. \cite{faulkner2013fresh} proposed a new version of this architecture for massive noisy sensors networks, and Cochran et al. \cite{cochran2009quake} presented an implementation using accelerometers connected to laptops and workstations, named QCN. Similarly, MyShake, proposed by Kong et al. \cite{kong2016myshake}, is a machine-learning-based EEW system that uses smartphones. The Earthquake Network (Finazzi et al. \cite{finazzi2016earthquake}) is a different research project that uses smartphones and spatial correlation to detect quakes. SeismoCloud (\cite{panizzi2016seismocloud}) is another earthquake early warning system built using smartphones and internet-of-things devices. All these systems differ from our proposal as they rely on a central system to collect all reports and make the final decision.

Another approach is the one described by CrowdQuake, from Huang et al. \cite{huang2020crowdquake}. CrowdQuake runs a CRNN on the fusion center, while smartphones at the edge collect samples of different lengths and stream them to the fusion center. While relying on the fusion center, this system differs from the previous ones because it can do both the decision and the detection in one step since the accuracy of the CRNN is very high. It also shows some architectural limits that we will describe in Section \ref{sec:stateofart}.

Fischer et al. in \cite{fischer2009model} described SOSEWIN, a self-organizing Earthquake Early Warning system using a wireless mesh network. They use a hybrid approach, where nodes act as local fusion centers. Instead, in our proposal, each node is independent of others.

Lee et al. \cite{lee2019smart} presented a custom-made board for EEW. The board contains common chipsets (like ESP8266) and custom software with an Artificial Neural Network for detection. They propose to send the alert to nearby smart devices (TV, smartphones) via low range transmissions (Bluetooth Low Energy) or Home Automation solutions for early warning alerts. Unlike other solutions, including ours, they do not use a network to send the alert to nearby houses; their alert is ``personal''.

\section{State of the art}
\label{sec:stateofart}

We present five different systems of crowdsensing EEW, which cover all the scenarios we can find in the literature. Other networks are similar to those presented here, so we do not cover them.

\subsection{Quake-catcher network}

QCN, Quake-Catcher Network, is an earthquake early warning system built by volunteers to ``fill the gap between the earthquake and traditional networks'' \cite{cochran2009quake}. It has been built over BOINC \cite{anderson2004boinc}. QCN uses MEMS sensors found in some laptop brands (usually in the anti-shock subsystem) and some USB accelerometer brands. According to \cite{cochran2009quake}, the sensitivity of these accelerometers is low, and the network is well suited for an earthquake of magnitude greater than 5.0.

QCN uses a Z-Score to detect potential quakes: when $z$ is above $3$, the sensor sends all relevant data to the fusion center, like max amplitude or timestamp. Then, the center will again use a Z-Score against the number of reports in a given area and time slice; a value of $z>6$ will trigger an EEW.

\subsection{MyShake}

MyShake\cite{kong2016myshake} is an earthquake early warning system developed by UC Berkeley Seismology Lab, designed to collect and process data on a smartphone and send possible quakes to a fusion center for confirmation. Volunteers can download a mobile application on their smartphone to join the network.

The MyShake mobile app reads the signal from the smartphone's internal MEMS accelerometer. Then it uses an artificial neural network to detect potential quakes and sends quake candidates to the fusion center, where a clustering algorithm reduces false positives.

\subsection{Crowdquake}

CrowdQuake, by Huang et al. \cite{huang2020crowdquake}, has a layered approach for earthquake detection. The lower layer, composed of dedicated smartphones or custom Internet-of-Things devices, senses the seismic data and streams it to an intermediate layer of ``gateways''. Each gateway is a GPU-equipped server that processes seismic samples from each sensor in a CRNN. Then, it sends data to a third and fourth layer for notification, monitoring, and visualization.

Differently from others, Crowdquake requires a stable and low latency network connection between sensors and gateways because samples from the accelerometers are sent to gateways (acting as fusion centers) for detection. This fact is a substantial limitation for the deployment, especially in remote sites.

\subsection{SOSEWIN}

SOSEWIN \cite{fischer2009model} is a decentralized wireless mesh network of sensors built using standard PC boards and external sensors. There is a hierarchy in the network between nodes, built using a leader election algorithm; the two most important kinds of nodes are the \textit{sensing} node and the \textit{leading} node. A leading node receives information from five sensing nodes and manages alerts from them and neighbors leading nodes. A sensing node filters the accelerometer sensor information with an IIR passband filter and some thresholds, using an internal state machine to refine the detection.

The leading node acts as a fusion center of a cluster of sensing nodes. Using leader election for leading nodes, SOSEWIN obtains High Availability.

\subsection{SeismoCloud}

In SeismoCloud \cite{panizzi2016seismocloud} smartphone apps and Internet-of-Things devices make the sensor network and connect to a fusion center. Both types of sensor nodes run an algorithm based on dynamic Z-Score for candidate quakes detection and send candidate quakes to the central server, where a clustering algorithm filters out false positives.

\section{Proposed architecture}

We propose a new architecture for EEW systems based on crowdsensing and the complete detection of earthquakes at the edge. The goal is to achieve fault tolerance using low-cost commodity hardware while enhancing the privacy and scalability of the system.

We will start describing each component of the system. Then we will draw a comprehensive picture of the architecture.

\subsection{Probes, Detectors, and Local Authorities}

There are two main roles in our network: the \textit{probe} and the \textit{detector}. A probe is a sensor capable of picking the acceleration signal and streaming it to the detector. The main role of the detector is to run the detection algorithm over the data stream from probes and match if there is any sign of an earthquake wave. One detector can link more probes, and we expect that, as the hardware for a probe is very cheap,  multiple probes will maximize the chance of detection.

In addition, the two roles can be assigned to the same device if it can read the accelerometer signal and run the detection algorithm simultaneously. For example, smartphones and some System-on-Chip boards have enough computing power to support such operations.

The detector is also equipped with a local alert device, like a PA speaker, to alert the owner of the detector itself. It is triggered either by a local earthquake alert or by a remote one.

The Local Authority is a central server that supports the network with non-critical services: it needs to keep detector registrations (for node discovery) and receive EEWs (for EEW logging). Both requirements can be implemented using a gossiping protocol between servers inside the local authority (thus implementing eventual consistency): this is possible due to the \textit{state-less} nature of these services.

\subsection{Network architecture}

\begin{figure}[htb]
    \centering
    \includegraphics[scale=0.4]{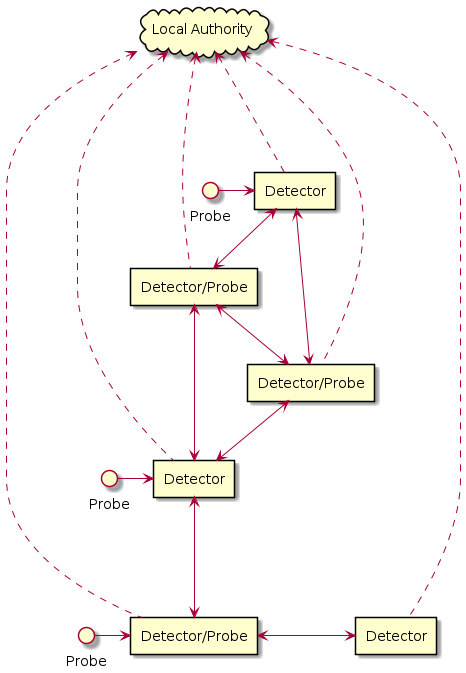}
    \caption{Example network. Sensors are linked to neighbors based on their own location}
    \label{fig:network}
\end{figure}

The network architecture that we propose is a decentralized mesh (Figure \ref{fig:network}). Each detector is connected to other neighbor detectors using direct peer-to-peer links. Probes connect to their nearest detector.

Detectors exchange EEW messages using a gossiping protocol: each message is forwarded to neighbor detectors until it reaches a certain distance from the reported quake location. Moreover, detectors report all quakes to the cloud service of the local authority to relay this information to other services (e.g., TV broadcast). 
Probes are not connected among them and do not participate in any message exchange between detectors.

\begin{table}[htb]
\begin{tabular}{ll}
\textbf{EEW message content} & \\ \hline
\textbf{Timestamp}       & Timestamp of the detection \\
\textbf{Origin location} & Location coordinates of the detector\\ & which originated the message \\
\textbf{Signal data}     & Signal window which triggered the warning
\end{tabular}
\caption{Earthquake Early Warning message content. This message is relayed between \textit{detectors}}
\label{tbl:msgcontent}
\end{table}

\subsection{Bootstrap sequence}
\label{sec:bootstrap}

\begin{figure}[htb]
    \centering
    \includegraphics[scale=0.49]{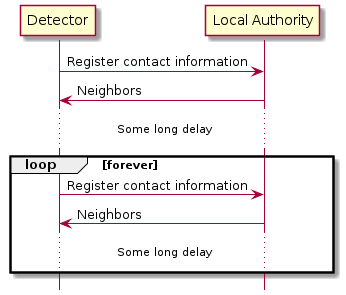}
    \caption{Detector bootstrap sequence diagram}
    \label{fig:detectorboot}
\end{figure}

When a detector powers up for the first time (Figure \ref{fig:detectorboot}), it starts a discovery phase of its neighbors using a service of registration and discovery of the local authority. The detector advertises its presence by using that service, and in turn, it receives the list of neighbors and details on how to connect to them.

After completing this exchange, the detector will connect to the indicated neighbors and keep those connections alive, ready to relay information about early warnings. Periodically, the detector repeats the registration and receives a new list of neighbors.

\begin{figure}[htb]
    \centering
    \includegraphics[scale=0.49]{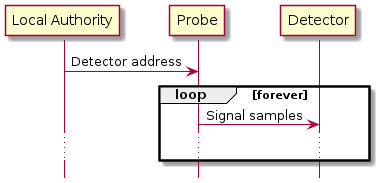}
    \caption{Probe bootstrap sequence diagram}
    \label{fig:probeboot}
\end{figure}

Differently, probes query the local authority for the detector that they should connect to. They do not advertise any information to the local authority (see Figure \ref{fig:probeboot}).

\subsection{Detection pipeline}

\begin{figure}[htb]
    \centering
    \includegraphics[scale=0.4]{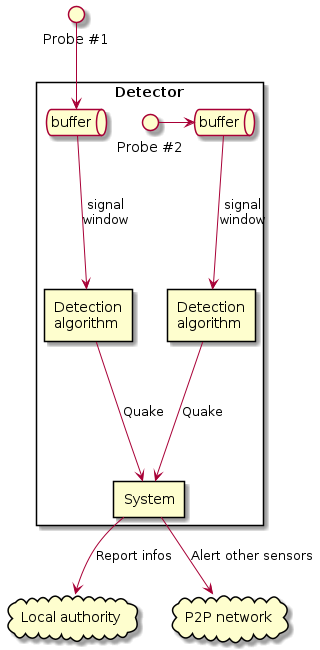}
    \caption{Pipeline diagram. Each probe is attached to its own buffer, which in turn feeds a dedicated instance of the detection algorithm. Probe \#3 is internal, as this detector has also the probe role}
    \label{fig:pipeline}
\end{figure}

Figure \ref{fig:pipeline} shows the detection pipeline. Probes stream the signal using their network connection to the detector. The detector has one buffer per probe, where it collects and stores the accelerometer signal for some time. At given intervals, a sliding signal window is extracted and sent to the detection algorithm.

If the detection algorithm detects a quake, the detector relays the earthquake alert, together with the signal itself and its location, to the local authority and neighbors, (Table \ref{tbl:msgcontent}).
The detector can release the alert according to a policy requiring most probes to report a quake.

\subsection{Scalability}

The architecture is capable of scaling indefinitely. No constraint limits the number of devices in the network. By employing a gossiping protocol between detectors, each message will reach a subset of the network, creating local, dynamic, and temporarily broadcast domains. By directly connecting sensors between them, there is no need to scale up a central server to handle sensors traffic.

\subsection{Fault tolerance}
\label{sec:fault}
This network is fully fault-tolerant. A fault of one sensor will not stop the gossiping. An EEW message can be prevented from reaching the entire network only if multiple faults occur so that the network temporarily splits into two or more partitions. However, the more sensors in the network, the less the chance of having such a split. Even if this split occurs, it will not affect messages from other sensors living inside other partitions. If sensors in the different partitions detect the quake, the EEW can still be sent to the whole network (albeit with different origins).

A fault on a specific sensor itself will not stop the detection: neighbors can still detect the earthquake, and they will still be able to pass information to others.

In case of faults in the local authority, which causes the unreachability of their service, new nodes will not be able to connect to the network. The local authority will not receive an EEW during the downtime. Already connected nodes will keep their current connections. Faults on the local authority service will not stop the EEW message gossiping at all.

\subsection{Privacy}

Accelerometer signal is fully processed locally on each detector. The local authority does not receive signals that are not classified as earthquakes by the detection algorithm itself. An attacker who wants to monitor sensors (for example, to recognize spoken words or detect the presence of people in a building) will need to attack specific sensors actively.

\section{Prototype implementation}

We present the following implementation as an example of the architecture presented above. This implementation is currently running in a test environment for our project.

\subsection{Sensors hardware}

The detector role device is a Raspberry Pi, made by the \textit{Raspberry Pi Foundation}. We tested both version 3 model B and version 4 (Table \ref{tbl:rpispecs}). It is a \textit{System-on-Chip} board with various ports (Ethernet, USB, HDMI, I$^2$C, GPIO), Wi-Fi, and Bluetooth wireless chipsets. For this prototype, we use the Ethernet port to provide an Internet connection to the detector, the I$^2$C bus to connect the accelerometer, and a Wi-Fi card to create a dedicated Wi-Fi network for external probes (in addition to the internal one).

\begin{table}[htb]
\begin{tabular}{lllll}
                    & \multicolumn{3}{c}{\textbf{Raspberry Pi}}                                                                                                                                                                                                        &                                                                     \\
\textbf{Model}    & \textbf{2B+}                                                                   & \multicolumn{1}{c}{\textbf{3B+}}                                               & \multicolumn{1}{c}{\textbf{4}}                                                 & \multicolumn{1}{c}{\textbf{NodeMCU}}                                \\ \hline
\textbf{CPU}      & \begin{tabular}[c]{@{}l@{}}BCM2836\\ Cortex-A7\\ 4-Core\\ 900 MHz\end{tabular} & \begin{tabular}[c]{@{}l@{}}BCM2837\\ Cortex-A53\\ 4-Core\\ 1.2GHz\end{tabular} & \begin{tabular}[c]{@{}l@{}}BCM2711\\ Cortex-A72\\ 4-Core\\ 1.5GHz\end{tabular} & \begin{tabular}[c]{@{}l@{}}106Micro\\ L106\\ @ 160 MHz\end{tabular} \\
\textbf{RAM}      & 1GB                                                                            & 1GB                                                                            & 4GB                                                                            & 128kBytes                                                           \\
\textbf{Disk}     & 64GB SD                                                                        & 64GB SD                                                                        & 64GB SD                                                                        & 4MBytes                                                             \\
\textbf{Wi-Fi}    & \begin{tabular}[c]{@{}l@{}}2.4 GHz\\ 5 GHz\end{tabular}                        & \begin{tabular}[c]{@{}l@{}}2.4 GHz\\ 5 GHz\end{tabular}                        & \begin{tabular}[c]{@{}l@{}}2.4 GHz\\ 5 GHz\end{tabular}                        & 2.4 GHz                                                             \\
\textbf{Ethernet} & Fast                                                                           & Fast                                                                           & Gigabit                                                                        &                                                                     \\
\textbf{GPIO}     & 40 pin                                                                         & 40 pin                                                                         & 40 pin                                                                         & 13 pin                                                             
\end{tabular}
\caption{Hardware specifications for prototype detectors and probes}
\label{tbl:rpispecs}
\end{table}

By connecting the MPU6050 directly to the Raspberry Pi,  probe and detector roles merge: the detector can read data directly from the accelerometer via the I$^2$C bus. The Raspberry Pi can also collect data from external probes.

Probe sensors use an MCU board and an accelerometer, packed to run on 5v from a power supply or battery. They transmit values using the Wi-Fi connection to the detector. The MCU board is the ``NodeMCU DEVKIT'' that contains a ESP8266 SoC \cite{kumar2015internet}, based on ESP-12 hardware. It has multiple GPIO ports and an integrated Wi-Fi network connection.

The accelerometer is the MPU6050, which is widely used in low-cost IoT applications involving acceleration measurements. It has been demonstrated by Crisnapati et al. \cite{8674321} and Lee et al. \cite{lee2019smart} that this accelerometer can be used in EEW applications. The MPU6050 provides a 100Hz feed via I$^2$C to the NodeMCU board or the Raspberry Pi board.

\subsection{Software}

The detector runs Raspberry PI OS (previously known as Raspbian), a Debian-like GNU/Linux distribution. Inside the OS, we loaded ``Podman'', an OCI image runner alternative to ``Docker'' that we chose as it is \textit{daemon-less}. The absence of a central process makes Podman more robust and less resource-hungry than Docker.

Podman runs multiple containers: the main container with the code we are presenting and the additional containers for external detection algorithms we are currently testing (machine learning engines). These additional containers are connected via gRPC to the main process. The use of containers simplifies the deployment of new algorithms for testing. While we built the main container executable using Go (and TensorFlow C bindings), additional containers were built using Python.

The probe runs a customized firmware that we built using the Expressif SDK for Arduino. The firmware reads data from the accelerometer sensor and sends the stream via WebSocket to the detector. The connection is using WebSocket to be compatible with HTTP middlewares, like network proxies and firewalls. We previously tested MQTT but found it too complex for this scenario. At probe boot, the firmware checks for updates and configuration, querying a central server. If it fails, it still connects to the detector.

\subsection{Detection pipeline}

\begin{figure}[htb]
    \centering
    \includegraphics[scale=0.49]{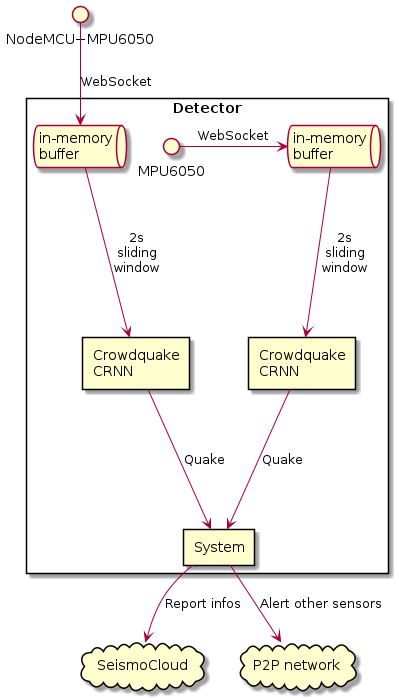}
    \caption{Prototype pipeline. The detection algorithm is the Crowdquake CRRN, and the time window is set to 2 seconds}
    \label{fig:implpipeline}
\end{figure}

The main container exposes a WebSocket endpoint for probes, and it reads the accelerometer connected to the GPIO of the Raspberry Pi. The code spawns multiple processes (based on the number of cores/CPUs of the Raspberry Pi) so that reading a local sensor while receiving a network stream does not interfere with each other.

The probe data stream is buffered in memory by the main container to have a 2-second signal window (200 values), with a 1-second sliding window, as shown in Figure \ref{fig:implpipeline}. The signal window is then sent to the detection algorithm, the Crowdquake CRNN in our prototype (running on the CPU).

The detection CRNN is run every second (as it receives a new set of samples each second), looking for quakes in the last 2 seconds of the signal. Each probe has its independent buffer, and CRNN is run in parallel on each buffer.

In our prototype setup, one CRNN is sufficient to trigger an earthquake early warning system.

\section{Results}

We ran our prototype in two different test environments: first, we tested the whole system using a single detector and feeding test samples to test the system's soundness. Then, we built an actual probe to test the speed of the detection for the whole system.

Our accuracy results for the detection algorithm matches Crowdquake one (presented in \cite{huang2020crowdquake}). We could cross-check values and results by sending the same dataset from Crowdquake via WebSocket to one detector, simulating an external probe. Checking the output of the whole detector is essential to assess that we maintain the same properties of the original detection algorithm.

The detection pipeline is capable of analyzing and output the result in few milliseconds, as shown in Table \ref{tbl:speed}. The detection latency for Raspberry Pi 4 is lower thanks to the faster CPU and different onboard bus wiring.

\begin{figure}[htb]
    \centering
    \includegraphics[scale=0.5]{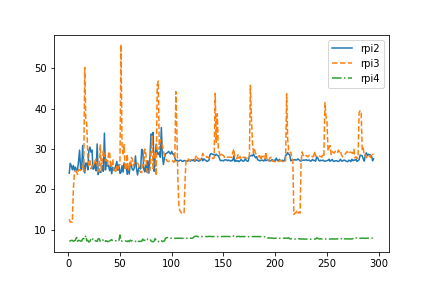}
    \caption{Detection latency per platform. Up-peaks are caused by the Go garbage collector}
    \label{fig:speed}
\end{figure}

The Go garbage collector is causing a spike that nearly doubles the detection latency when it executes concurrently to the detection algorithm (see Figure \ref{fig:speed}). It can be further optimized by running the garbage collector manually, rewriting the buffer code (where most of the allocation takes place), or switching to a language with no automatic memory management (e.g., Rust, C).

The detector code loads the network in memory, and it launches an ``empty'' run to pre-fill the system cache (this is the reason why in Figure \ref{fig:speed} the first value is in the same range of values of the rest of the plot).

\begin{table}[htb]
\centering
\begin{tabular}{llll}
                                 \textbf{Raspberry Pi} & \multicolumn{1}{c}{\textbf{\begin{tabular}[c]{@{}c@{}}Average\\ time\end{tabular}}} & \multicolumn{1}{c}{\textbf{\begin{tabular}[c]{@{}c@{}}Standard\\ deviation\end{tabular}}} & \multicolumn{1}{c}{\textbf{90-percentile}} \\ \cline{2-4} 
\multicolumn{1}{l|}{\textbf{2B}} & 27.19 ms                                                                            & 1.61 ms                                                                                   & 28.73 ms                                   \\
\multicolumn{1}{l|}{\textbf{3+}} & 27.78 ms                                                                            & 5.36 ms                                                                                   & 30.74 ms                                   \\
\multicolumn{1}{l|}{\textbf{4}} & 7.84 ms                                                                             & 0.41 ms                                                                                   & 8.37 ms                                   
\end{tabular}
\caption{CRNN response speed for 2-second signal (200 values), averages on 300 samples}
\label{tbl:speed}
\end{table}

\subsection{Limits}

In this work, we did not address the security of the network. This area has multiple aspects, mainly related to the trust in early warnings from neighbors' sensors. Today, any byzantine probe in the network can cause a false alarm by sending an alert with a signal that resembles an earthquake wave (downloadable from public datasets). A byzantine detector can even send an earthquake early warning. We originally designed the protocol message so that a signal could be sent together with the EEW as a primary security measure: we planned to check that the signal attached to the EEW was triggering an EEW by replicating the detection on each detector. However, we did not clearly define or implement this part at this time, so we omitted this from the proposal, and it constitutes future work.

Another limit is that we did not address the problem of setting up a secure transmission between peers. In the prototype, we implemented a plain-text protocol in which attackers can eavesdrop on the message exchange (loss of confidentiality) and even inject or modify messages. However, this problem can be solved trivially by using widely studied and deployed protocols like TLS \cite{tlsrfc}.

It is worth underlining that the accuracy of the detection algorithm plays a central role in the trust of this system. Users will trust an EEW system with this architecture only if they receive very low false positives and false negatives EEWs. We are working on this by allowing different detection algorithms to plug in to compare their accuracy.

\section{Conclusions and future work}

We described a crowdsensing EEW architecture that moves the computation to the edge, with detector nodes that probe the environment and process information from nearby probes to detect earthquakes locally. Our approach tolerates multiple node faults and partial network disruption and keeps all data locally, enhancing privacy. We described our proposal's rationale, explained its architecture, and presented an implementation using Raspberry, NodeMCU, and the Crowdquake machine learning model.

On-going research on this topic focuses on the security of this architecture and its implementations. It is essential to find a viable and secure solution to the problem of trust in \textit{peer-to-peer} EEW message exchange to use this architecture in crowdsensing networks. At least the system should resist some byzantine nodes.

Another focus of our current research is the implementation of the architecture that we presented over low-power long-range radio protocols (LPWAN), such as LoRa. These wireless protocols are very effective in both long-range transmissions and power efficiency compared to Wi-Fi networks. However, they usually lack coordination, so collision-avoidance algorithms like water-filling cannot be used (unlike in LoRaWAN, where water-filling can be implemented in BTS\cite{cuomo2017explora} or in the control plane\cite{cuomo2018adaptive}), and we will need to overcome this limitation. The detection network can be deployed seamlessly from big cities to remote sites using LPWAN for IoT, LTE for smartphones, and FWA/FTTx for others. In the big cities, the Internet is ubiquitous, while in remote sites, LPWAN can be a low-cost, low-latency alternative to satellite links.

\bibliographystyle{plain}
\bibliography{biblio}

\end{document}